\pdfoutput=1

\documentclass{appolb}
\usepackage{amssymb}
\usepackage{hyperref}
\usepackage{graphicx}
\usepackage{amsmath}
\usepackage{epsfig,bm,slashed}

\usepackage[numbers,sectionbib,sort&compress]{natbib}

\begin{document}

\title{Wigner and Husimi partonic distributions of the pion \\ in a chiral quark model}

\author{Wojciech Broniowski$^{1,2}$\thanks{Wojciech.Broniowski@ifj.edu.pl}, Enrique Ruiz Arriola$^3$\thanks{earriola@ugr.es}
\address{$^{1}$Institute of Physics, Jan Kochanowski University, PL-25406~Kielce, Poland}
\address{$^{2}$The H. Niewodnicza\'nski Institute of Nuclear Physics, \\ Polish Academy of Sciences, PL-31342~Cracow, Poland}
\address{$^{3}$Departamento de F\'isica At\'omica, Molecular y Nuclear \\ and Instituto Carlos I de F\'{\i}sica Te\'orica y Computacional,
                   Universidad de Granada, E-18071 Granada, Spain}
}

\maketitle

\begin{abstract}
Generalized transverse momentum distributions (GTMDs), the Wigner, and
the Husimi distributions of quarks in the pion are evaluated in a
chiral quark model at the one-loop-level. Analytic expressions are
obtained for GTMDs, allowing for a qualitative discussion of their
features, whereas the Wigner and the Husimi distribution are obtained
with numerical integration of simple formulas. We explain the features
of the Wigner distributions, in particular their non-positivity. In
our model, the Husimi distributions, which are interpreted as
coarse-grained Wigner distributions, are not mathematically
positive-definite, but the magnitude of their negative values is tiny
and occurs at large transverse momenta and impact parameters. Hence,
as expected, coarse-graining leads to better behaved functions from
the point of view of the probabilistic interpretation.
\end{abstract}

\maketitle

\section{Introduction}

In this paper we provide a nontrivial model example of the
Wigner~\cite{Ji:2003ak} and Husimi~\cite{Hagiwara:2014iya} quark
distributions in the pion at a low non-perturbative quark-model energy
scale~\cite{Broniowski:2007si}, confronting the issue of positivity of
these distributions.  Various kinds of partonic distributions are
related to each other by integrations or by Fourier transforms (for
the complete ``genealogical tree'', see, e.g., Fig.~2
of \cite{Diehl:2015uka}), in particular the Wigner distribution is a
Fourier transform of the forward generalized transverse momentum
distribution (GTMD) (see also
\cite{Belitsky:2005qn,Lorce:2011ni,Lorce:2011kd,Mukherjee:2014nya,Chakrabarti:2017teq,More:2017zqq}).
Naturally, positivity of partonic distributions is a necessary feature
for their probabilistic interpretation. As a matter of fact, it is
highly nontrivial, and requesting that both (the forward) GTMD and the
Wigner distributions are positive definite relates to a difficult
mathematical problem of having both a function and its Fourier
transform positive definite~\cite{Giraud:2005vx}.

Even in Quantum Mechanics, the Wigner
distributions~\cite{PhysRev.40.749} 
are in general not positive definite.  A way of curing this problem with appropriate Gaussian coarse
graining was proposed by Husimi~\cite{1940264}. The construction leads
to a positive-definite phase space density.
A few years ago the concept of the Husimi distributions was introduced
to the partonic physics by Hagiwara, Hatta, and
Ueda~\cite{Hagiwara:2014iya,Hagiwara:2016kam}. Using an example of the light-front quark model wave functions from~\cite{Mukherjee:2014nya},
these authors have shown that the corresponding non-positive Wigner distribution is made positive via the Husimi coarse
graining.

In this paper we work in a covariant one-quark-loop framework of low
energy quantum field theoretical chiral quark models, complying to all
the Lorentz and gauge covariance requirements. As a result, both the
charge and momentum sum rules are satisfied.  The model, supplemented
with the QCD evolution, was found to reasonably reproduce wide-ranging
properties of the pion, such as the parton distribution functions
(PDFs)~\cite{Davidson:1994uv,Davidson:2001cc,Weigel:1999pc}, the
distribution amplitude~\cite{RuizArriola:2002bp}, the generalized
distribution functions (GPDs)~\cite{Broniowski:2007si}, the
generalized form factors~\cite{Broniowski:2008hx}, the quasi
distribution amplitude~\cite{Broniowski:2017wbr}, the quasi and pseudo
PDFs, GPDs, and Ioffe-time
distributions~\cite{Broniowski:2017gfp,Shastry:2022obb}, or the double
distribution functions~\cite{Broniowski:2019rmu}.

Chiral quark models are generically intended to model hadron structure
at a low resolution scale in a non-perturbative scheme where explicit
gluonic degrees of freedom are absent.  All results of this paper
pertain to a low non-perturbative quark model scale
$\sim \!\!330$~MeV~\cite{Broniowski:2007si}, which is defined as the
scale where quarks are the only degrees of freedom.  We do not perform
the DGLAP evolution to higher scales, whereby gluons would be radiatively
generated, as this is highly nontrivial for GTMDs or its Fourier transforms.

We find that in our model both the Wigner and the Husimi distributions are not
positive definite. However, the negative values in the Husimi
distributions are very small compared to the Wigner distributions, so the
problem, though present mathematically, is from a ``practical'' point
of view significantly improved through the Husimi coarse graining.

The Wigner distributions in the light-front approach were recently
studied in~\cite{Liu:2015eqa,Ma:2018ysi,Kaur:2019kpi,Ahmady:2020ynt},
where the non-positivity feature is also manifest.

\section{Definitions}

We denote a generic four-vector as $a=(a^0,\bm{a},a^3)$, with the bold
face indicating the transverse part $\bm{a}=(a^1,a^2)$. We also use
$a_T\equiv |\bm{a}|$. The light-cone coordinates are $a^\pm= (a^0 \pm
a^3)/\sqrt{2}$, whence $a \cdot b= a^+ b^-+a^- b^+
- \bf{a}\cdot\bm{b}$. The symmetric kinematic convention (the Breit
frame) is used, where the momenta of the initial and final pions are,
correspondingly, $p\pm \tfrac{1}{2} {\Delta}$. We also introduce a
null vector $n$, whence
\begin{eqnarray} 
p=(p^0,\bm{0},0), \;\;\; \Delta=(0,\bm{\Delta},0), \;\;\; n=\frac{1}{p^0}(1,\bm{0},-1). \label{eq:kin}
\end{eqnarray}
The initial and final pions are on mass shell, hence
\begin{eqnarray} 
p^0 = \sqrt{m_\pi^2 +\tfrac{1}{4} \bm{\Delta}^2}, \label{eq:onmass}
\end{eqnarray}
where $m_\pi$ is the pion mass. 
Covariantly, 
\begin{eqnarray} 
n^2=0, \; p^2=m_\pi^2 - \tfrac{1}{4} t, \; t=-\bm{\Delta}^2, \; p\cdot n=1, \; p \cdot \Delta = 0, \; \xi \equiv n \cdot \Delta = 0, \label{eq:kincov}
\end{eqnarray}
where the last condition reflects the vanishing skewness $\xi$ for the considered forward case. In the partonic framework
\begin{eqnarray}
k \cdot n = x, \label{eq:defx}
\end{eqnarray}
where in the so-called symmetric convention $k$ is the average (before and after the interaction) momentum of the probed quark (cf.~Fig.~\ref{fig:feyndia}).

The leading-twist chirally even generalized transverse-momentum distributions (GTMDs) at zero skewness
are defined via the following matrix elements,
\begin{eqnarray}
&& \delta_{ab}\delta_{\alpha\beta}{\cal G}^{I=0}(x,\bm{\Delta},\bm{k})+i\epsilon^{abc}\tau^c_{\alpha\beta}{\cal G}^{I=1}(x,\bm{\Delta},\bm{k}) = 
\frac{1}{2}\!\int \! \frac{d z^-d^2\bm{z}}{(2\pi)^3} e^{i x \, p^+\! z^- \!-i \bm{k}\cdot \bm{z}} \nonumber \\
&& ~~~~  \times \left . \langle\pi^b(p-\tfrac{1}{2}\bm{\Delta})|\overline{\psi}_\alpha(-\tfrac{z}{2}) \gamma^+ {\cal L} \,
\psi_\beta(\tfrac{z}{2})|\pi^a((p+\tfrac{1}{2}\bm{\Delta})\rangle \right |_{z^+=0}, \label{eq:gtmd}
\end{eqnarray}
where the superscripts in ${\cal G}$ indicate the isospin combinations, $\psi$ stands for the quark field, indices $\alpha$ and $\beta$ 
represent the quark flavor, $a$, $b$ denote the isospin of the pions, while $c$ is the isospin of the probing operator. 
The symbol ${\cal L}$, that makes the expression gauge invariant, is the staple-shaped~\cite{Collins:2003fm} Wilson 
line extending along the light-cone coordinate $z^-$.
Definition similar to~(\ref{eq:gtmd}) holds also for the gluons, not considered here.

The good isospin combinations of GTMDs from Eq.~(\ref{eq:gtmd}) are related to GTMDs of quarks and antiquarks as follows:
\begin{eqnarray}
{\cal G}^{q,\bar{q}}(x,\bm{\Delta},\bm{k})=\label{eq:Hq} \frac{1}{2} \left [ {\cal G}^{I=0}(x,\bm{\Delta},\bm{k})\pm{\cal G}^{I=1}(x,\bm{\Delta},\bm{k})\right ]. \label{eq:qq}
\end{eqnarray} 
By general arguments of the Lorentz covariance, the function ${\cal G}^{q}(x,\bm{\Delta},\bm{k})$ has the support $x \in [0,1]$, 
whereas ${\cal G}^{\bar{q}}(x,\bm{\Delta},\bm{k})$ has the support $x \in [-1,0]$. It complies to the convention that 
\begin{eqnarray}
&& \int d^2\bm{k} \, {\cal G}^{q}(x,\bm{\Delta}=0,\bm{k}) = q(x), \label{eq:qq1} \\
&& \int d^2\bm{k}\, {\cal G}^{\bar{q}}(-x,\bm{\Delta}=0,\bm{k}) = - \bar{q}(x), \nonumber
\end{eqnarray}
where $q(x)$ and $\bar{q}(x)$ are the (positive) parton distribution functions (PDFs) of, correspondingly, quarks and antiquarks with the support $x\in[0,1]$.

\section{GTMDs in chiral quark models at the one-loop level \label{sec:GTMDs}}

For simplicity, from now on we work in the strict chiral limit of the vanishing current quark mass.
\begin{figure}[tb]
\centering
\includegraphics[height=0.3\textwidth]{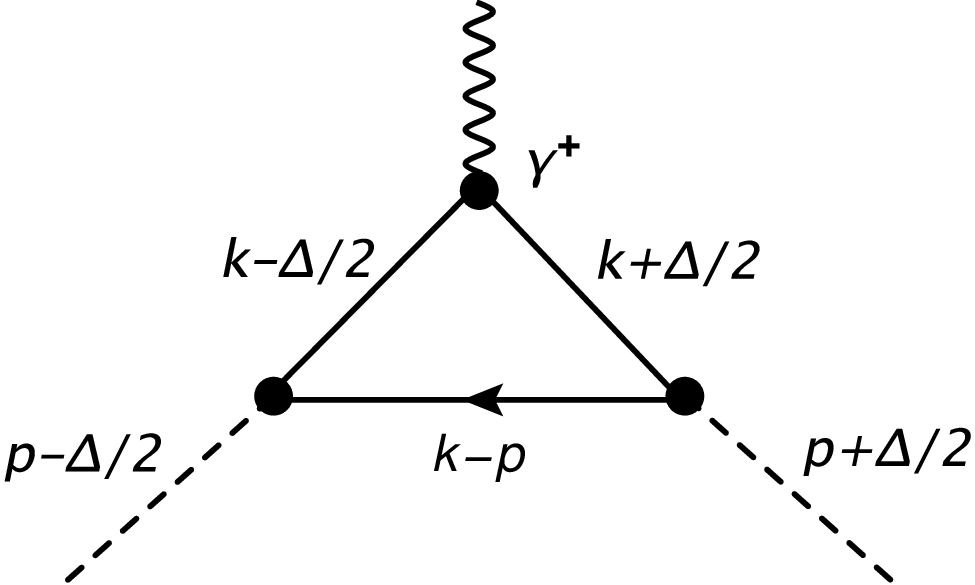} 
\caption{One-quark-loop diagram for the evaluations of the quark forward  GTMDs in chiral quark models. Condition~(\ref{eq:defx}) is understood and the
transverse momentum $\bm{k}$ is fixed, hence the loop integration is only over $k^-$. The dashed lines indicate the pion states and the solid line is the 
quark propagator. The distribution of antiquarks is given by the crossed diagram. \label{fig:feyndia}}
\end{figure}
The model used in this paper is a nonlinear realization of the chiral quark model, with the Lagrangian density 
\begin{eqnarray}
{\cal L}(x)=\bar{\psi}(x) \left [ i \slashed{\partial} - \omega \, e^{-i \gamma_5  \tau^a \phi^a(x)} \right ] \psi(x). \label{eq:lag}
\end{eqnarray}
Here $\omega$ denotes the (constituent) quark mass following from the dynamical chiral symmetry breaking,
$\tau_a$ are the Pauli isospin matrices, and $\phi^a$ is the pion field, which 
is a Goldstone boson according to the Nambu--Jona-Lasinio mechanism. The pion decay constant $f=86$~MeV in the chiral limit. 
With the nonlinear realization, one maintains chiral symmetry without introducing the $\sigma$ field of the linear model, which results in simpler results
for the off-forward case (not studied here). From Eq.~(\ref{eq:lag})
we read out the quark-pion vertex \mbox{$i {\omega}/{f} \, \gamma_5  \tau^a$}.

At the one-quark-loop level, which corresponds to the leading-$N_c$ quark-model calculation, GTMDs are evaluated from the diagram of Fig.~\ref{fig:feyndia}. 
From definition~(\ref{eq:gtmd}) transformed into the momentum space, the Feynman rules yield the following expression,
\begin{eqnarray}
&&{\cal G}^q(x,\bm{\Delta},\bm{k})=  \frac{N_c \omega^2}{f^2} \int \frac{dk^+ dk^-}{(2\pi)^4} \delta(k^+\!-x) 
{\rm Tr}\left [ \gamma^+ S_{k-\tfrac{q}{2}} \gamma_5 S_{k-p} \gamma_5  S_{k+\tfrac{q}{2}} \right ], \nonumber \\ \label{eq:tr}
\end{eqnarray}
where 
\begin{eqnarray}
S_k=\frac{i}{\slashed{k}-\omega +i \epsilon}
\end{eqnarray}
denotes the quark propagator and the trace is over the Dirac space. 
The evaluation of the trace and the standard one-loop reduction yields the basic structure
\begin{eqnarray}
{\cal G}^{q}(x,\bm{\Delta},\bm{k})= \tfrac{1}{2} [I_{-} +I_{+}+ (1-x)J ]. \label{eq:IKq} 
\end{eqnarray}
The one-loop functions appearing above are defined as
\begin{eqnarray}
&& I_{s} = -i \frac{N_c \omega^2}{4 \pi^4 f^2}\int dk^+dk^- \frac{\delta(k\cdot n - x)}{[(k- p)^2 -\omega^2+i \epsilon][(k+s \tfrac{q}{2})^2 -\omega^2+i \epsilon]}, \nonumber \\
&& J = -i \frac{N_c \omega^2 t}{4 \pi^4 f^2}\int dk^+dk^- \times \label{eq:bubbles}\\
&& ~~~~~ \frac{\delta(k\cdot n - x)}{[(k-p)^2 -\omega^2+i \epsilon][(k+\tfrac{q}{2})^2 -\omega^2+i \epsilon][(k-\tfrac{q}{2})^2 -\omega^2+i \epsilon]}, \nonumber 
\end{eqnarray}
where  $s = \pm 1$. Note that the definition of $J$ 
includes the factor of $t$. The arguments of all the functions in the above equations are $(x,\bm{\Delta},\bm{k})$.  
However, the distributive structure of Eq.~(\ref{eq:IKq}) is generic and holds also for the Wigner and the Husimi distributions discussed later on, or for the GPDs as derived 
in~\cite{Broniowski:2007si}. The loop functions $I_s$ and $J$ are evaluated explicitly in Appendix~\ref{app:loops}. 

\section{Regularization}

The evaluation of GPDs, which are integrals of GTMDs over $d^2 \bm{k}$, requires regularization, 
since the two-point function $I_s$ is logarithmically divergent. However, the need for regularization is physically motivated also for finite quantities, in order to 
separate the hard momenta, not treated in low-energy models, and the soft momenta, crucial for the dynamics. 
The spectral regularization~\cite{RuizArriola:2003bs} used in this paper, or the Pauli-Villars 
regularization, may be carried out in an elegant way on the formulas for the loop functions. 
The key feature here is that the product of the Klein-Gordon propagators written in the Schwinger representation,
appearing in the one-loop functions such as (\ref{eq:bubbles}) (cf.~Appendix~\ref{app:loops}), contains generically the factor 
$\exp[-(\alpha_1 + \alpha_2+...+\alpha_n)\omega^2)]$, where $\alpha_i$ are the Schwinger parameters. Then, regularizations involving 
distributions of the quark mass $\omega$ lead to simple (analytic) expressions.\footnote{Note that such a prescription is equivalent to 
subtractions, hence it may promptly affect the positivity property of the parton distributions.}

The spectral regularization~\cite{RuizArriola:2003bs,Broniowski:2007si} amounts to the evaluation of the quark loop integrals
according to the prescription
\begin{align}
    \mathcal{A}^{\rm SQM} = \int_C d\omega  \rho(\omega) \mathcal{A}
\end{align}
where, $\mathcal{A}$ sands for an unregularized amplitude, $\rho(\omega)$ is a properly chosen spectral density function, 
and $C$ is a contour of integration in the complex $\omega$ plane (cf. Fig.~1 in \cite{RuizArriola:2003bs}).
In SQM, one has a possibility to implement exactly the vector meson dominance in the pion electromagnetic form factor, 
which is successful phenomenologically. More details are provided in Appendix~\ref{app:sqm}.

\section{Quark GTMD} 

With the formulas from Appendix~\ref{app:loops} it is straightforward to write down the explicit formulas for the forward quark GTMD in SQM, 
\begin{eqnarray}
{\cal G}^{q}(x,\bm{\Delta},\bm{k})&=&\frac{3 M_V^3}{\pi} 
\left[\frac{1}{\left(4 \bm{k}_+^2+M_V^2\right)^{5/2}}+\frac{1}{\left(4\bm{k}_-^2+M_V^2\right)^{5/2}}\right] \label{eq:tgpdexp} \\
&+& \frac{3 M_V^3 (1-x) \bm{\Delta}^2}{2\pi \bm{k}\cdot \bm{\Delta}} \left[\frac{1}{\left(4 \bm{k}_+^2+M_V^2\right)^{5/2}}-\frac{1}{\left(4\bm{k}_-^2+M_V^2\right)^{5/2}}\right] , \nonumber
\end{eqnarray}
with the shorthand notation
\begin{eqnarray}
\bm{k}_\pm= \bm{k}\pm \tfrac{1}{2}(1-x) \bm{\Delta}. \label{eq:kpm}
\end{eqnarray}
The form of ${\cal G}^{q}$ exhibits explicitly the scaling
\begin{eqnarray}
{\cal G}^{q}(x,\bm{\Delta},\bm{k})={\cal G}^{q}[(1-x)\bm{\Delta},\bm{k}], \label{eq:scaleG}
\end{eqnarray} 
where the arguments in the right-hand side indicate that ${\cal G}^{q}$ is a function of the combination $(1-x)\bm{\Delta}$ (and $\bm{k}$) only.
This feature is specific to the one-loop model in the chiral limit.

\begin{figure}[tb]
\centering
\includegraphics[width=0.65\textwidth]{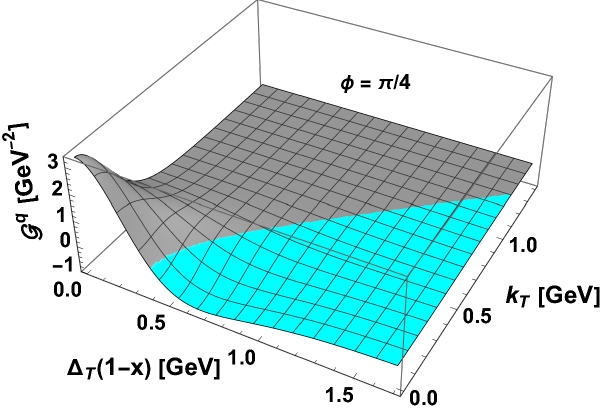} 
\caption{Quark forward ($\xi=0$) GTMD of the pion in SQM, plotted as a function of the quark transverse momentum $k_T$ and the combination
 $(1-x) \Delta_T$ for a sample value of the angle $\phi$  between $\bm{k}$ and $\bm{\Delta}$. The lighter (blue) color indicates 
 negative values of the distribution. \label{fig:gtmd}}
\end{figure}

In Fig.~\ref{fig:gtmd} we plot  ${\cal G}^{q}$  as a function of $k_T$ and $(1-x) \Delta_T$ for a selected
value of the angle between $\bm{k}$ and $\bm{\Delta}$, namely $\phi=\pi/4$ (for other values of $\phi$ the results are qualitatively similar). We note that 
${\cal G}^{q}$ is positive near the origin, but at larger vales of $\Delta_T$ it assumes negative values, indicated with a lighter (blue) color.

For the special case of $\bm{\Delta=0}$ Eq.~(\ref{eq:tgpdexp}) reduces to the $k_T$-unintegrated PDF and equals to
\begin{eqnarray}
{\cal G}^{q}(x,\bm{0},\bm{k})=\frac{6 M_V^3}{\pi  \left(4 k^2+M_V^2\right)^{5/2}},\label{eq:qkT} 
\end{eqnarray}
in agreement with~\cite{Broniowski:2003rp}. The forward ($\xi=0$) GPD is equal to~\cite{RuizArriola:2003bs}
\begin{eqnarray}
{G}^{q}(x,\bm{\Delta})=\int d^2 \bm{k}\,\, {\cal G}^{q}(x,\bm{\Delta},\bm{k}) = \frac{M_V^2 \left(M_V^2+t (x-1)^2\right)}{\left(M_V^2-t (x-1)^2\right){}^2}, 
\end{eqnarray}
and the corresponding forward impact parameter distribution 
\begin{eqnarray}
 q (x , \bm{b}) &=& \int \frac{d^2 \bm{\Delta}}{(2\pi)^2} \, e^{-i \bm{\Delta} \cdot \bm{b}} G^q (x , \bm{\Delta} ) \nonumber \\ 
 &=& \frac{M_V^2}{2\pi (1-x)^2} \left[ \frac{ b_T M_V }{1-x}
K_1 \left( \frac{ b_T M_V }{1-x} \right) - K_0 \left( \frac{ b_T M_V }{1-x} \right) \right] \, , 
 \label{eq:qb}
\end{eqnarray}
where $K_0$ and $K_1$ are modified Bessel functions (we correct a
global minus sign typo of~\cite{Broniowski:2003rp}). This
function is {\it not} positive definite at small values of $b_T \le
(1-x) /M_V z_0$, with $z_0\simeq 0.596$ fulfilling $K_0 (z_0)= z_0 K_1 (z_0)$, and actually at $b_T \to 0$ diverges logarithmically.

The electromagnetic and gravitational form factors are~\cite{Broniowski:2008hx}
\begin{eqnarray}
&& F_V(t)=\int_0^1 dx\, {G}^{q}(x,\bm{\Delta})=\frac{M_V^2}{M_V^2-t}, \\
&& \theta_2(t)= 2 \int_0^1 dx\, x\, {G}^{q}(x,\bm{\Delta}) = \frac{M_V^2}{t} \log \left(\frac{M_V^2}{M_V^2-t}\right). \nonumber
\end{eqnarray}
Clearly, $F_V(0)=\theta_2(0)=1$, expressing the charge charge and momentum (mass) sum rules.

\begin{figure}[tb]
\centering
\includegraphics[width=0.62\textwidth]{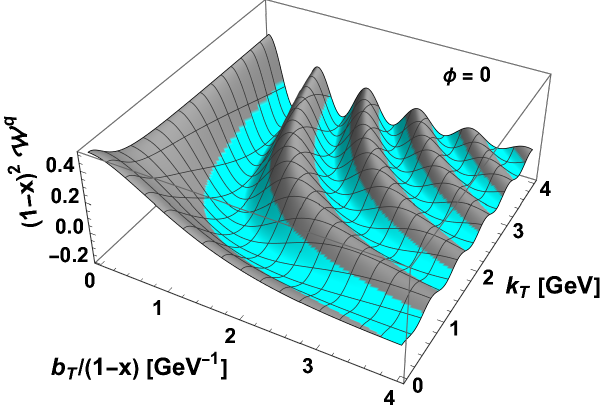} \\ \includegraphics[width=0.62\textwidth]{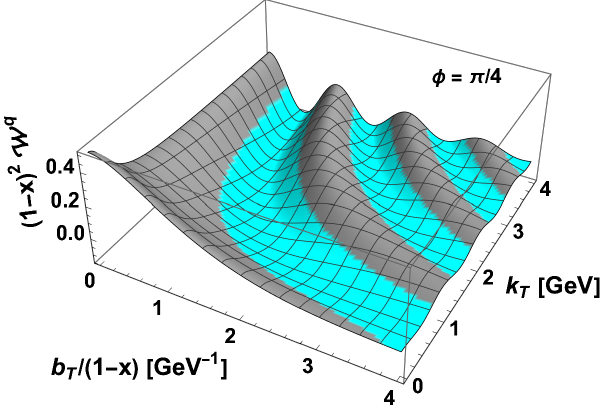}  \\ \includegraphics[width=0.62\textwidth]{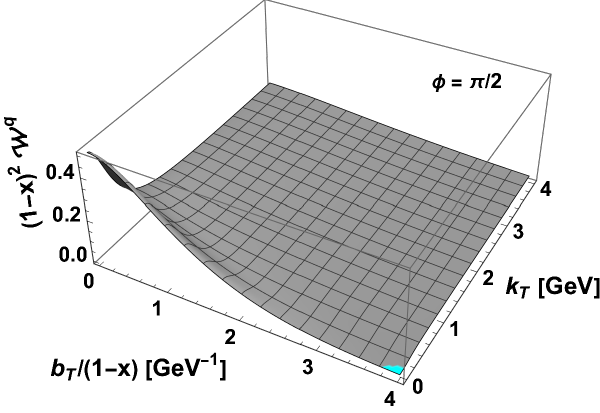} 
\caption{The quark Wigner distribution of the pion in SQM multiplied with $(1-x)^2$, plotted as a function of the quark transverse momentum 
$k_T$ and the combination
 $b_T/(1-x)$ for three sample values of the angle $\phi$  between $\bm{b}$ and $\bm{\Delta}$. The lighter (blue) color indicates 
 negative values of the distributions. \label{fig:wig}}
\end{figure}

\section{Wigner distributions}

The Wigner distribution is the Fourier transform of the forward GTMD from the  momentum space, $\bm{\Delta}$, into the impact parameter space, $\bm{b}$,
\begin{eqnarray}
{\cal W}^q(x,\bm{b},\bm{k})=\int \frac{d^2 \bm{\Delta}}{(2\pi)^2} e^{-i \bm{\Delta}\cdot \bm{b} }\,{\cal G}^{q}(x,\bm{\Delta},\bm{k}). \label{eq:wigdef}
\end{eqnarray}
The marginal distributions are the impact-parameter distribution ,
\begin{eqnarray}
\int d^2 \bm{k} \,{\cal W}^q(x,\bm{b},\bm{k})= \, q(x, \bm{b}), \label{eq:b-dist}
\end{eqnarray}
and the TMD distribution,
\begin{eqnarray}
\int  d^2 \bm{b} \,{\cal W}^q(x,\bm{b},\bm{k})=\, T^q (x,\bm{k} ) = {\cal G}^q(x,\bm{0},\bm{k}). \label{eq:k-distribution}
\end{eqnarray}
The normalization from the double integration yields the quark PDF,
\begin{eqnarray}
\int d^2 \bm{b}\, d^2 \bm{k} \,{\cal W}^q(x,\bm{b},\bm{k})=\int d^2 \bm{k} \,{\cal G}^q(x,\bm{0},\bm{k})=q(x). \label{eq:Wnorm}
\end{eqnarray}
From Eq.~(\ref{eq:scaleG}) it follows that in our model ${\cal W}^{q}$ obeys the scaling
\begin{eqnarray}
{\cal W}^{q}(x,\bm{b},\bm{k})= \frac{1}{(1-x)^2} {\cal W}^{q}[\bm{b}/(1-x),\bm{k}]. \label{eq:scaleW}
\end{eqnarray}

With Eqs.~(\ref{eq:a1},\ref{eq:a2}) we get the following semi-analytic formula in SQM,
\begin{eqnarray}
(1-x)^2 {\cal W}^q(x,\bm{b},\bm{k}) &=& \frac{1}{\pi^2} e^{-B_T M_V} (B_T M_V+1) \cos(2 B_T k_T \cos \phi) \label{eq:wigint}\\
&-& \left[ \frac{\partial^2}{\partial b_T^2} + \frac{1}{b_T} \frac{\partial }{\partial b_T}\right ] \int_{-1}^1 dz \, 
\frac{M_V^3}{2\pi^2 u^5}  e^{-\sqrt{2} B_T u} \times \nonumber \\
&& ~~~[B_T u (2 B_T u+3 \sqrt{2})+3] \cos(2 B_T k_T z \cos \phi), \nonumber
\end{eqnarray}
where 
\begin{eqnarray}
&& u=\sqrt{M_V^2 + 4 k_T^2 (1-z^2)}, \label{eq:BT} \;\; B_T=\frac{b_T}{1-x}. \nonumber
\end{eqnarray}
The first line in Eq.~(\ref{eq:wigint}) originates form the two-point functions, whereas the integral over the parameter $z$ comes from the three-point functions
specified in Appendix~\ref{app:loops}. The appropriate differentiation with respect to $b_T$ brings down the factor of $\Delta^2$ present in 
the integrand of the definition of the Wigner transform~(\ref{eq:wigdef}).
The transverse momentum and impact parameter marginal distributions are checked to be given by Eq.~(\ref{eq:qkT}) and  Eq.~(\ref{eq:qb}) respectively.

The quark Wigner distribution of the pion obtained in our model is plotted in Fig.~\ref{fig:wig}. We clearly note an oscillatory character, following directly from the 
form of the argument of the cosine functions in Eq.~(\ref{eq:wigint}). The contribution of the two-point functions $I_s$ contains $\cos(2 B_T k_T \cos \phi)$, 
hence its zeros are at hyperbolas in the $b_T-k_T$ plane at locations 
\begin{eqnarray}
b_T k_T= \frac{(2j+1)\pi}{4 \cos \phi}  (1-x), \;\;\; j\in \mathbb{Z}. \label{eq:bk}
\end{eqnarray}
The contribution of the three-point loop function $J$ contains an integral of \mbox{$\cos(2 z B_T k_T \cos \phi)$}, with the rest of the integrand sharply peaked at $z=\pm 1$, which results in a condition close to Eq.~(\ref{eq:bk}). The behavior of Eq.~(\ref{eq:bk}) is clearly seen in Fig.~\ref{fig:wig}. In particular, we 
note that the period of the oscillations increases as $1/\cos \phi$, and no oscillations occur at the boundaries $b_T=0$ or $k_T=0$, or  
when $\bm{b}$ and $\bm{k}$ are perpendicular, i.e., $\phi=\pi/2$.

\section{Husimi distributions}

\begin{figure}[tb]
\centering
\includegraphics[width=0.62\textwidth]{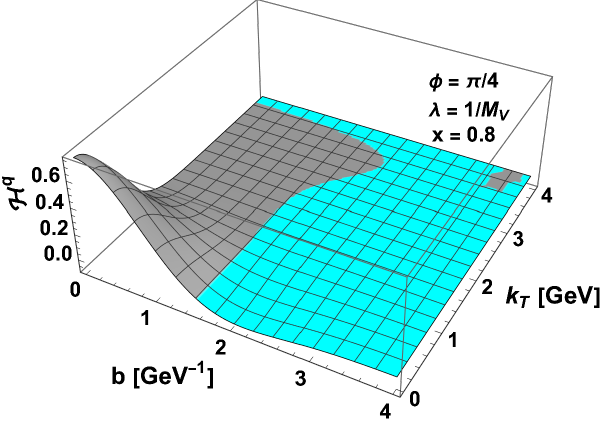} 
\caption{The Husimi distributions of quarks in the pion in SQM, 
plotted as a function of the quark transverse momentum $k_T$ and the transverse coordinate
 $b_T$ for the angle $\phi=\pi/4$  between $\bm{b}$ and $\bm{\Delta}$, $x=0.8$, and a sample value of the smearing scale $\lambda=1/M_V$. 
 The lighter color indicates negative values of the distributions. \label{fig:hus8}}
\end{figure}

Consider the smeared distribution
\begin{eqnarray}
{\cal H}^q(x,\bm{b},\bm{k}) &=& \frac{\Lambda^2}{\pi^2\lambda^2} \int d^2\bm{b}'\, d^2\bm{k}'\, 
e^{-(\bm{b} - \bm{b}')^2/\lambda^2 - (\bm{k} - \bm{k}')^2 \Lambda^2}{\cal W}^q(x,\bm{b}',\bm{k}'), \nonumber \\ \label{eq:husdef}
\end{eqnarray}
where $\lambda$ and $\Lambda$ are the smearing parameters (with dimension of length).
In Quantum Mechanics, the choice $\Lambda=\lambda$ warrants the positivity of~(\ref{eq:husdef}), which then becomes the celebrated Husimi distribution. 
Therefore we
follow~\cite{Hagiwara:2014iya,Hagiwara:2016kam} and fix $\Lambda=\lambda$ in our field-theoretic model as well. The normalization is 
\begin{eqnarray}
\int d^2 \bm{b}\, d^2 \bm{k} \, {\cal H}^q(x,\bm{b},\bm{k})=q(x). \label{eq:Hnorm}
\end{eqnarray}
Note that ${\cal H}^q$ does not exhibit scaling analogous to
Eq.~(\ref{eq:scaleW}) (unless we also scaled $\lambda$ with
$(1-x)$ and kept a separate $\Lambda$ unscaled, which we do not do). Since the
scale in our model is the vector meson mass $M_V$, we show the case
$\lambda=M_V$, while other values lead to qualitatively similar
results.
The procedure of obtaining the Husimi distributions in our model is outlined in Appendix~\ref{app:loops}.

\begin{figure}[tb]
\centering
\includegraphics[width=0.62\textwidth]{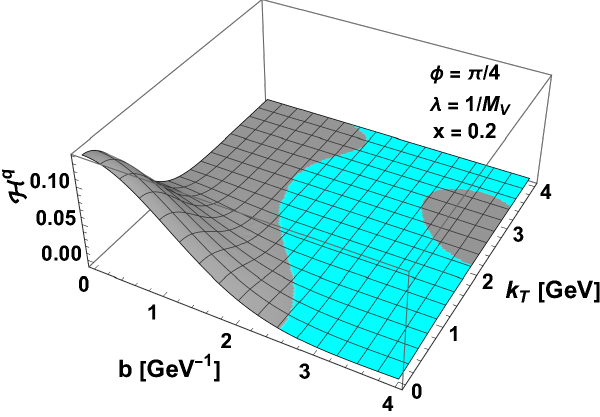} 
\caption{Same as in Fig.~\ref{fig:hus8}, but for $x=0.2$. \label{fig:hus2}}
\end{figure}

The results for $x=0.8$ and $x=0.2$ are plotted, correspondingly, in
Figs.~\ref{fig:hus8} and \ref{fig:hus2}. We take $\phi=\pi/4$, as the
results for other angles are qualitatively similar. We notice that in
SQM, the Husimi distributions are not strictly positive, as at larger
values of $b_T$ ($b_T \gtrsim 2~{\rm GeV}^{-1}\simeq 0.5~{\rm fm}$) we
find broad regions with negative values.  However, contrary to the
case of the Wigner distributions of Fig.~\ref{fig:wig}, the
distributions in these regions are very shallow compared to the value
of the function at the origin. One might say that the positivity
breaking has been cured to a large extent by the application of the
Husimi coarse-graining procedure, although mathematically the problem
does persist. The results are qualitatively similar at various values
of $x$, as can be seen by comparing Figs.~\ref{fig:hus8}
and \ref{fig:hus2}.

The problem of the lack of strict positivity of the Husimi distributions in our field-theoretic model may possibly
be traced back 
to the subtractive nature of regularization, which also led to  the non-positivity of the marginal impact parameter
distribution of Eq.~(\ref{eq:qb}). We note that a similar situation was also encountered in the impact-parameter behavior of
the double parton distributions~\cite{Broniowski:2019rmu}, where an analogous
conflict between regularization and positivity was faced. 
Further studies of this intriguing and genuine field theoretic issue
of regularization vs. positivity are left for a future research.

\section{Conclusions}

Several merits of the model study presented in this paper should be
underlined. First, the model is simple but non-trivial, leading to a
rich set of results for various pion properties. In particular, the
framework allows to obtain the quark GTMD of the pion
analytically. This GTMD complies to all the formal requirements, in
particular the resulting GPDs possess polynomiality, with the charge
and momentum sum rules satisfied, although no factorization in the
kinematic variables is present in the model.  The corresponding Wigner
distribution contains one numerical integration, and the Husimi
distribution -- two, with simple integrands.  In the obtained Wigner
distribution, we can clearly see the origin and the pattern of the
oscillations leading to the breaking of positivity.  In the Husimi
distributions, these oscillations are smoothed by coarse-graining.
Although the Husimi distributions in our model are not strictly
positive, the negative values are tiny and appear at larger values of
the impact parameter, $b_T \gtrsim 0.5 ~{\rm fm}$.

\medskip

We thank Yoshitaka Hatta and Krzysztof Golec-Biernat for useful comments.
WB acknowledges the support by the Polish National
Science Centre (NCN) grant 2018/31/B/ST2/01022  and ERA by  project PID2020-114767GBI00 
funded by MCIN/AEI/10.13039/501100011033 and Junta de
Andaluc\'{\i}a grant FQM-225. 

\begin{appendix}  

\section{Loop functions} \label{app:loops}

In this Appendix we work with Euclidean momenta, corresponding to their Minkowski counterparts, but denoted with the same symbols.
An effective way~\cite{Broniowski:2007si} to evaluate the one-loop integrals with a momentum constraint is to use the Schwinger representation of the Klein-Gordon propagator, 
\begin{eqnarray}
\frac{1}{k^2+\omega^2}=\int_0^\infty\!\!\! d\alpha \, e^{-\alpha (k^2+\omega^2)}. \label{eq:sch}
\end{eqnarray}
The two-point function from Eq.~(\ref{eq:bubbles}) is then expressed as
\begin{eqnarray}
  && \hspace{-3mm}  I_s =  \frac{4 N_c \omega^2}{f^2}\!\! \int \!\frac{dk_0 dk_3}{(2\pi)^2} \!\! \int\!\frac{d\xi}{2\pi}
    e^{i\xi(k\cdot n-x)}\!\!\int_0^\infty \!\!\!\!\!\! d\alpha \int_0^\infty \!\!\!\!\!\! d\beta \, e^{-\alpha[(k-p)^2+\omega^2]-\beta[(k+s \Delta /2)^2+\omega^2]}\nonumber\\
  && \;\; = \frac{N_c}{4 \pi ^3 f^2}\!\! \int_0^\infty \!\!\!\!\!\! d\alpha \int_0^\infty \!\!\!\!\!\! d\beta \, 
  e^{-\frac{(2 \bm{k} (\alpha+\beta)+\beta s \bm{\Delta} )^2}{4 (\alpha+\beta)}-(\alpha+\beta)\omega^2} \delta[x(\alpha+\beta)-\alpha],
\end{eqnarray}
where the intermediate steps follow exactly~\cite{Broniowski:2007si}.
In SQM, with Eq.~(\ref{eq:st}), the procedure yields 
\begin{eqnarray}
  && \hspace{-3mm}  I_s =  \frac{M_V^3}{4 \pi ^{3/2}}\!\! \int_0^\infty \!\!\!\!\!\! d\alpha \int_0^\infty \!\!\!\!\!\! d\beta \, (\alpha+\beta)^{3/2} 
  e^{-\frac{(2 \bm{k} (\alpha+\beta)+\beta s \bm{\Delta} )^2}{4 (\alpha+\beta)}-\tfrac{1}{4}(\alpha+\beta)M_V^2} \delta[x(\alpha+\beta)-\alpha] \nonumber \\
  && \;\;= \frac{6 M_V^3}{\pi  \left([2\bm{k}+  s (1-x)\bm{\Delta}]^2+M_V^2\right)^{5/2}}. \label{eq:is}
\end{eqnarray}

For the three-point function from Eq.~(\ref{eq:bubbles}) an analogous calculation yields
\begin{eqnarray}
  && \hspace{-3mm}  J =  -\frac{4 N_c \omega^2 t}{f^2}\!\! \int \!\frac{dk_0 dk_3}{(2\pi)^2} \!\! \int\!\frac{d\xi}{2\pi}
    e^{i\xi(k\cdot n-x)}\!\!\int_0^\infty \!\!\!\!\!\! d\alpha \int_0^\infty \!\!\!\!\!\! d\beta \int_0^\infty \!\!\!\!\!\! d\gamma \times  \\
  && \hspace{2cm} \, e^{-\alpha[(k-p)^2+\omega^2]-\beta[(k+ \Delta /2)^2+\omega^2]-\gamma[(k -\Delta /2)^2+\omega^2]}\nonumber\\
  && \;=- \frac{N_c  \omega^2 t}{4 \pi ^3 f^2} \!\!\int_0^\infty \!\!\!\! dh\, h \int_0^1 \!\!\! dy \int_0^{1} \!\!\! dz \,  \theta(1-y-z)\delta(1\!-\!y\!-\!z\!-\!x) \times\nonumber \\
  &&\hspace{2cm}  e^{-h [\bm{k}^2+\omega^2+\bm{\Delta}\cdot\bm{k} (y-z)+\frac{1}{4} \bm{\Delta}^2 (y+z)^2]}, \nonumber
  \end{eqnarray}
where $h=\alpha+\beta+\gamma$, $y=\beta/h$, and $z=\gamma/h$. In SQM
\begin{eqnarray}
  && \hspace{-3mm}  J = - \frac{M_V^3 t}{4 \pi^{3/2}} \!\!\int_0^\infty \!\!\!\! dh\, h^{5/2} \int_0^1 \!\!\! dy \int_0^{1} \!\!\! dz \,  \theta(1-y-z)\delta(1\!-\!y\!-\!z\!-\!x) \times\nonumber \\
  &&\hspace{2cm}  e^{-h [{\bm{k}}^2+\frac{1}{4} M_V^2+{\bm{\Delta}}\cdot\bm{k} (y-z)+\frac{1}{4} {\bm{\Delta}}^2 (y+z)^2]}  \nonumber \\
  && \;\;= \frac{3 M_V^3 \bm{\Delta}^2}{\pi \bm{k}\cdot \bm{\Delta}} \left[\frac{1}{\left(4 \bm{k}_-^2+M_V^2\right)^{5/2}}-\frac{1}{\left(4\bm{k}_+^2+M_V^2\right)^{5/2}}\right]
  \end{eqnarray}
in the notation~(\ref{eq:kpm}) used in the last line.  

The Fourier transform of $I_s$ is elementary,
\begin{eqnarray}
&& \int \frac{d^2\bm{\Delta}}{(2\pi)^2} e^{-i \bm{\Delta}\cdot \bm{b} }\,I_s(x,\bm{\Delta},\bm{k}) = \int \frac{d^2\bm{\Delta'}}{(2\pi)^2}
\frac{6 M_V^3 e^{-i \bm{b}\cdot \left( \bm{\Delta'}-\frac{2 s \bm{k}}{1-x}\right)}}{\pi  \left[M_V^2+\bm{\Delta'}^2 (1-x)^2\right]^{5/2}} \nonumber \\
&& ~~~  =\frac{B_T {M_V}+1}{\pi^2 (1-x)^2} e^{-B_T (M_V-2 i s k_T)} , \label{eq:a1}
\end{eqnarray}
with $\bm{\Delta'}=\bm{\Delta}+2s\bm{k}/(1-x)$ and the notation of Eq.~(\ref{eq:BT}) used. 
For the case of the three-point function, it is useful to express it via the integral representation
\begin{eqnarray}
J/\bm{\Delta}^2=\frac{16 M_V^3}{\pi^{3/2}} \int_0^\infty \!\!\! dh \int_{-1}^1 \!\! dz \, h^{5/2} 
    e^{-h ( 4 \bm{k}^2 - 4 z \bm{k}\cdot \bm{\Delta} + \bm{\Delta}^2 + M_V^2)} \label{eq:Jr}
\end{eqnarray}
(for simplification, here we get rid of the factor $\bm{\Delta}^2$, which is later restored via differentiation with respect to $\bm{b}$).
Carrying out first the Fourier transform from $\bm{\Delta}$ to $\bm{b}$, and then integrating over $h$ yields the formula 
\begin{eqnarray}
&& (1-x)^2 \int \frac{d^2\bm{\Delta}}{(2\pi)^2} J/\bm{\Delta}^2 = \int_{-1}^1 dz \, 
\frac{M_V^3}{\pi^2 u^5}  e^{-\sqrt{2} B_T u} \times \label{eq:a2}\\
&&\hspace{2cm} [B_T u (2 B_T u+3 \sqrt{2})+3] \cos(2 B_T k_T z \cos \phi), \nonumber 
\end{eqnarray}
in the notation of Eq.~(\ref{eq:BT}). The one-dimensional integral over $z$ is left to be done numerically.

To evaluate the Fourier transforms needed for the Husimi distributions, it is convenient to rewrite Eq.~(\ref{eq:is}) in the integral representation 
\begin{eqnarray}
    I_s= \frac{4 M_V^3}{\pi^{3/2}} \int_0^\infty dh\, h^{3/2} e^{-h (4 k^2_s+ M_V^2)},  \;\;s=\pm1.
\end{eqnarray}
Then the Gaussian integrations over $\bm{b}'$, $\bm{k}'$, and $\bm{q}$ are carried out in a straightforward way, leaving a numerical integration over $h$. 
In the case of the three-point function $J$, a completely analogous procedure is carried out on expression~(\ref{eq:Jr}), with the integrals over the $h$ and $z$ 
remaining as numerical.
   
\section{Spectral regularization} \label{app:sqm}

The spectral function is~\cite{RuizArriola:2003bs}
\begin{eqnarray}
    \rho(\omega) = \frac{1}{2\pi i}\frac{1}{\omega}\frac{1}{\left(1-4\omega^2/M_V^2 \right)^{5/2}}, \label{eq:rho}
\end{eqnarray}
and the closed quark line is associated with the integral over a suitably chosen contour $C$. The form~(\ref{eq:rho}) implements
vector meson dominance in the pion electromagnetic form factor. For the present applications we need the formula
\begin{eqnarray}
\int_C d\omega\, \omega^2  \rho(\omega) e^{-A \omega} = \frac{A^{3/2} M_V^5}{24 \sqrt{\pi }} e^{-\frac{A M_V^2}{4}}, \;\; A \ge 0. \label{eq:st}
\end{eqnarray}
In SQM there is the following relation between the pion decay constant and the vector meson mass:
\begin{eqnarray}
f^2=\frac{N_c M_V^2}{24 \pi^2}. \label{eq:fMv}
\end{eqnarray}

\end{appendix}  

\bibliography{Ref}

\end{document}